\pdfoutput=1
\documentclass[sigconf, table, rgb, natbib=false, screen, 9pt, nonacm]{acmart}

\usepackage{subcaption}
\usepackage{newtxmath}
\usepackage[textsize=tiny]{todonotes}
\usepackage{nicefrac}
\usepackage{csquotes}
\usepackage{mathtools}
\usepackage{enumitem}
\usepackage[binary-units=true, detect-all=true, round-precision=2,
    table-format = 1.2,
    table-number-alignment=right]{siunitx}
\usepackage{url}
\usepackage{graphicx}
\usepackage{setspace}
\usepackage{multirow, makecell}
\usepackage{array}
\usepackage{marvosym}
\usepackage{xcolor}
\usepackage{verbatim}
\usepackage{nicematrix}
\usepackage{csvsimple}
\let\Bbbk\relax %
\usepackage{zx-calculus}
\usepackage{soul}
\usepackage{etoolbox}
\makeatletter
\patchcmd{\SOUL@ulunderline}{\dimen@}{\SOUL@dimen}{}{}
\patchcmd{\SOUL@ulunderline}{\dimen@}{\SOUL@dimen}{}{}
\patchcmd{\SOUL@ulunderline}{\dimen@}{\SOUL@dimen}{}{}
\newdimen\SOUL@dimen
\makeatother

\usepackage[style=ieee, maxnames=6, minnames=1, date=year, doi=false,isbn=false,backend=biber, sortcites, dashed=false]{biblatex}
\AtEveryBibitem{%
  \clearname{editor}%
  \clearfield{series}%
  \clearfield{isbn}%
  \clearfield{issn}%
  \clearfield{volume}%
  \clearfield{number}%
  \clearfield{pages}%
  \clearfield{urldate}%
  \clearfield{url}%
  \clearfield{urldate}%
}
\renewcommand*{\bibfont}{\footnotesize}
\makeatletter
\let\oldtheequation\theequation
\renewcommand\tagform@[1]{\maketag@@@{\ignorespaces#1\unskip\@@italiccorr}}
\renewcommand\theequation{(\oldtheequation)}
\makeatother

\tikzset{%
        terminal/.style={draw,rectangle,inner sep=2pt,font=\footnotesize,very thick},
        zeronode/.style={fill, draw, circle, minimum width=2pt, inner sep=0pt,color=black},
        qubit/.style={draw,circle,inner sep=0pt,minimum width=0.4cm,minimum height=0.4cm,font=\footnotesize,color=black, thin},
        edgeOne/.style={color=omega0,ultra thick},
        edgeMOne/.style={color=omega4,ultra thick},
        edgeSqrt/.style={color=omega0},
        edgeMSqrt/.style={color=omega4},
        edgeSqrt3/.style={color=omega0,thin},
        edgeOmega0/.style={color=omega0,ultra thick},
        edgeOmega1/.style={color=omega1,ultra thick},
        edgeOmega2/.style={color=omega2,ultra thick},
        edgeOmega3/.style={color=omega3,ultra thick},
        edgeOmega4/.style={color=omega4,ultra thick},
        edgeOmega5/.style={color=omega5,ultra thick},
        edgeOmega6/.style={color=omega6,ultra thick},
        edgeOmega7/.style={color=omega7,ultra thick},
        edge0/.style={edge from parent path={(\tikzparentnode) to[out=-130, in=90] (\tikzchildnode)}},
        edge1/.style={edge from parent path={(\tikzparentnode) to[out=-100, in=90] (\tikzchildnode)}},
        edge2/.style={edge from parent path={(\tikzparentnode) to[out=-80, in=90] (\tikzchildnode)}},
        edge3/.style={edge from parent path={(\tikzparentnode) to[out=-50, in=90] (\tikzchildnode)}},
      }

      \tikzset{%
        font={\footnotesize},
        vertex/.style={draw,circle,inner sep=0pt,minimum width=0.5cm,minimum height=0.5cm},
        zeroterm/.style={below,inner sep=0pt,font=\tiny}
}

\newcommand{\qop}[1]{\ensuremath{\mathit{#1}}}
\newtheorem{example}{Example}
\newtheorem{theorem}{Theorem}[section]
\newtheorem{lemma}[theorem]{Lemma}
\def\exampleautorefname{Example}
\renewcommand{\figureautorefname}{Fig.}
\renewcommand{\sectionautorefname}{Section}
\renewcommand{\subsectionautorefname}{Section}
\renewcommand{\equationautorefname}{Eq.}
\usepackage{tikz}
\usetikzlibrary{backgrounds,fit,decorations.pathreplacing,calc,matrix,arrows.meta,positioning,decorations.pathreplacing,quantikz}
\authorsaddresses{}

\renewcommand{\baselinestretch}{0.99}

\emergencystretch=\maxdimen
\begin{document}
\title{Equivalence Checking of Parameterized Quantum Circuits}
\subtitle{Verifying the Compilation of Variational Quantum Algorithms}
\author{Tom Peham}
\affiliation{%
        \institution{Chair for Design Automation, Technical University of Munich, Germany}
}
\email{tom.peham@tum.de}

\author{Lukas Burgholzer}
\affiliation{%
        \institution{Institute for Integrated Circuits, Johannes Kepler University Linz, Austria}
}
\email{lukas.burgolzer@jku.at}

\author{Robert Wille}
\affiliation{%
        \institution{Chair for Design Automation, Technical University of Munich, Germany}
}
\affiliation{%
        \institution{Software Competence Center Hagenberg GmbH, Austria}
}
\email{robert.wille@tum.de}
\begin{abstract}
  Variational quantum algorithms have been introduced as a promising class of quantum-classical hybrid algorithms that can already be used with the noisy quantum computing hardware available today by employing \emph{parameterized quantum circuits}.
  Considering the \mbox{non-trivial} nature of quantum circuit compilation and the subtleties of quantum computing, it is essential to verify that these parameterized circuits have been compiled correctly.
  Established equivalence checking procedures that handle \mbox{parameter-free} circuits already exist.
  However, no methodology capable of handling circuits with parameters has been proposed yet.
  This work fills this gap by showing that verifying the equivalence of parameterized circuits can be achieved in a purely symbolic fashion using an equivalence checking approach based on the \mbox{ZX-calculus}. At the same time, proofs of inequality can be efficiently obtained with conventional methods by taking advantage of the degrees of freedom inherent to parameterized circuits.
  We implemented the corresponding methods and proved that the resulting methodology is complete.
  Experimental evaluations (using the \emph{entire} parametric ansatz circuit library provided by Qiskit as benchmarks) demonstrate the efficacy of the proposed approach. The implementation is open source and publicly available as part of the equivalence checking tool QCEC~(\mbox{\url{https://github.com/cda-tum/qcec}}) which is part of the Munich Quantum Toolkit (MQT).
\end{abstract}
\maketitle

\section{Introduction}\label{sec:introduction}

Quantum computers promise a computational advantage over classical computers for certain problems~\cite{aruteQuantumSupremacyUsing2019,zhongQuantumComputationalAdvantage2020,huangProvablyEfficientMachine2022}.
Far-term quantum algorithms such as Shor's algorithm for integer factorization~\cite{shorPolynomialtimeAlgorithmsPrime1997} and Grover search~\cite{groverFastQuantumMechanical1996} require robust, large-scale error correction schemes to retrieve a meaningful result from their execution~\cite{devittQuantumErrorCorrection2013}.
\emph{Noisy Intermediate Scale Quantum} (NISQ) computers~\cite{preskillQuantumComputingNISQ2018}, the ones that are available today, are not able to implement this error correction yet.
Due to the high gate error rates and short coherence times of NISQ devices, the depth of any quantum circuit to be run on them is inherently limited.

\emph{Variational Quantum Algorithms}~\cite{cerezoVariationalQuantumAlgorithms2020} have been proposed to achieve a computational advantage despite these limitations.
These algorithms use quantum programs as subroutines in a classical optimization routine.
The optimization loop iteratively improves the parameters of a quantum circuit \emph{ansatz}, which, in turn, is used to estimate some desired quantity, such as the expectation value of some observable.
This general variational framework has been applied to solve problems in chemistry~\cite{mcardleQuantumComputationalChemistry2020}, finance~\cite{eggerQuantumComputingFinance2020}, discrete optimization~\cite{hadfieldQuantumApproximateOptimization2019}, and more.

But, before running an ansatz on a target device, a compilation step has to be performed where the circuit is translated into a natively supported gate-set and routed such that it conforms to the device's topological constraints---a costly and non-trivial procedure~\cite{liTacklingQubitMapping2019,willeMappingQuantumCircuits2019,boteaComplexityQuantumCircuit2018,tanOptimalLayoutSynthesis2020}.
To avoid the necessity of recompiling a quantum circuit in each iteration step of a variational algorithm, the circuit is usually compiled once in \emph{parameterized} form in which the parameters tuned by the classical optimization routine are not bound to specific values.
This compiled parameterized circuit can then be used in the optimization loop without having to perform all compilation steps over and over again.

The increasing use of parameterized circuits in the development of quantum algorithms has also brought along the need for verifying that these circuits have been compiled correctly.
Established equivalence checking methods such as proposed in~\cite{burgholzerVerifyingResultsIBM2020,amyLargescaleFunctionalVerification2019,burgholzerAdvancedEquivalenceChecking2021,kissingerReducingTcountZXcalculus2020,chun-yuAccurateBDDbasedUnitary2022} are able to prove that a compiled circuit still adheres to its specification after compilation. However, these approaches cannot handle parameterized circuits.
The only way to verify the compilation of variational quantum algorithms with these methods is to check the equivalence of the original and the instantiated compiled ansatz in each iteration.
Since equivalence checking is a hard problem for even one instance---it is, in fact, a QMA-complete problem~\cite{janzingNonidentityCheckQMAcomplete2005}---solving it over and over again is hardly a feasible approach.

In this work, we propose, for the first time, a methodology to solve the equivalence checking problem for parameterized quantum circuits.
The proposed, multi-stage method starts out by using an approach based on the \mbox{ZX-calculus}~\cite{vandeweteringZXcalculusWorkingQuantum2020,coeckePicturingQuantumProcesses2018} to try and prove the equivalence of both parameterized circuits.
If this does not succeed, parameters of the circuits are instantiated, and a conventional, complete equivalence checker is employed. In order to make this check as easy as possible, we derive an instantiation scheme that allows simplifying most of the parameterized gates from the circuits. Furthermore, we prove this yields a complete equivalence checking procedure for parameterized quantum circuits.

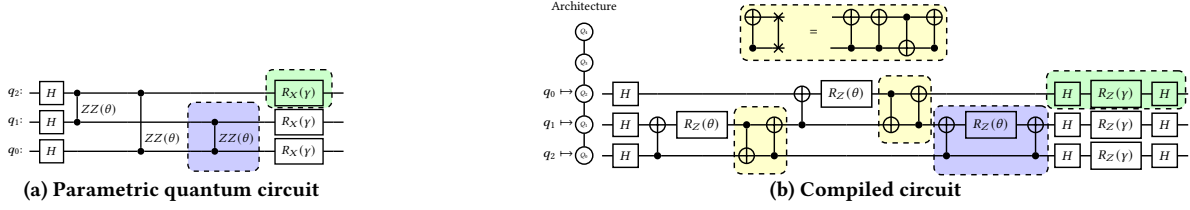
\begin{figure*}[t]
  \centering

  \begin{subfigure}[b]{0.33\linewidth}
  \centering
    \resizebox{.8\linewidth}{!}{
      \begin{quantikz}[column sep=6pt, row sep={0.6cm,between origins}]
        \lstick{$q_2$:} & \gate{H} & \ctrl{1}                                               & \qw & \ctrl{2}                                               & \qw & \qw & \qw                                                                                                                  & \qw & \gate[style={fill=green!20}]{R_X(\gamma)}\gategroup[1,steps=1,style={dashed,rounded corners,fill=green!20, inner xsep=2pt,yshift=0.09cm, inner ysep=.5pt},background]{} & \qw& \qw \\
\lstick{$q_1$:} & \gate{H} & \control[label={[label distance=.1cm]60:ZZ(\theta)}]{} & \qw & \qw                                                    & \qw & \qw & \ctrl{1}\gategroup[2,steps=2,style={dashed,rounded corners,fill=blue!20, inner xsep=2pt,inner ysep=1pt,xshift=-0.3cm},background]{} & \qw & \gate{R_X(\gamma)}                                                                                                              & \qw& \qw \\
\lstick{$q_0$:} & \gate{H} & \qw                                                    & \qw & \control[label={[label distance=.1cm]60:ZZ(\theta)}]{} & \qw & \qw & \control[label={[label distance=.1cm]60:ZZ(\theta)}]{}                                                               & \qw & \gate{R_X(\gamma)}                                                                                                              & \qw& \qw
\end{quantikz}
            }\vspace*{-1.3mm}
            \caption{Parametric quantum circuit}\label{fig:qaoa}
  \end{subfigure}\hspace{.039\linewidth}
  \begin{subfigure}[b]{0.62\linewidth}
  \centering
\resizebox{.8\linewidth}{!}{
      \begin{tikzpicture}[qnode/.style={circle, draw=black,scale=0.5},thick]
    \node[qnode] (Q0) at (-6,-0.4) {$Q_0$};
    \node[qnode] (Q1) at (-6, 0.2) {$Q_1$} ;
    \node[qnode] (Q2) at (-6 , 0.8) {$Q_2$};
    \node[qnode] (Q3) at (-6, 1.4) {$Q_3$};
    \node[qnode] (Q4) at (-6, 2) {$Q_4$};
    \draw[] (Q0.north) -- (Q1.south);
    \draw[] (Q1.north) -- (Q2.south);
    \draw[] (Q2.north) -- (Q3.south);
    \draw[] (Q3.north) -- (Q4.south);

    \node[] (q0) at (-6.5,-0.4) {$q_2 \mapsto$};
    \node[] (q1) at (-6.5, 0.2) {$q_1 \mapsto$} ;
    \node[] (q2) at (-6.5 , 0.8) {$q_0 \mapsto$};

    \node[]  at (-6 , 2.5) {Architecture};
      \node[] at(-1, 2.1) {
    \begin{quantikz}[column sep=6pt, row sep={0.6cm,between origins}]
      \targ{}\gategroup[2,steps=7,style={dashed,rounded corners,fill=yellow!20, inner xsep=0.01pt,inner ysep=0.01pt},background]{} & \swap{1} & \midstick[2,brackets=none]{$=$} & \targ{}   & \targ{}   & \ctrl{1} & \targ{}   \\
       \ctrl{-1}                                                                                                                   & \targX{} &                                 & \ctrl{-1} & \ctrl{-1} & \targ{}  & \ctrl{-1}
    \end{quantikz}
  };
    \node[] at (0,.35) {
          \begin{quantikz}[column sep=6pt, row sep={0.6cm,between origins}]
      \qw & \gate{H} & \qw       & \qw                & \qw      & \qw       & \targ{}   & \gate{R_Z(\theta)} & \ctrl{1}\gategroup[2,steps=2,style={dashed,rounded corners,fill=yellow!20, inner xsep=0.01pt,inner ysep=0.01pt},background]{} & \targ{}   & \qw                           & \qw                                      & \qw                           & \gate[style={fill=green!20}]{H}\gategroup[1,steps=3,style={dashed,rounded corners,fill=green!20, inner xsep=2pt, inner ysep=.5pt,yshift=0.08cm},background]{} & \gate[style={fill=green!20}]{R_Z(\gamma)} & \gate[style={fill=green!20}]{H} & \qw \\
      \qw & \gate{H} & \targ{}   & \gate{R_Z(\theta)} & \ctrl{1}\gategroup[2,steps=2,style={dashed,rounded corners,fill=yellow!20, inner xsep=0.01pt,inner ysep=0.01pt},background]{} & \targ{}   & \ctrl{-1} & \qw                & \targ{}  & \ctrl{-1} & \targ[style={fill=blue!20}]{}\gategroup[2,steps=3,style={dashed,rounded corners,fill=blue!20, inner xsep=0.1pt,inner ysep=.5pt},background]{} & \gate[style={fill=blue!20}]{R_Z(\theta)} & \targ[style={fill=blue!20}]{} & \gate{H} & \gate{R_Z(\gamma)} & \gate{H} & \qw \\
      \qw & \gate{H} & \ctrl{-1} & \qw                & \targ{}  & \ctrl{-1} & \qw       & \qw                & \qw      & \qw       & \ctrl{-1}                     & \qw                                      & \ctrl{-1}                     & \gate{H} & \gate{R_Z(\gamma)} & \gate{H} & \qw
    \end{quantikz}
  };

\end{tikzpicture}}\vspace*{-2.5mm}
\caption{Compiled circuit}\label{fig:qaoa-compiled}
\end{subfigure}

\vspace*{-3mm}
\caption{Compiling a parametric circuit to a device}
  \label{fig:running-example}
  \vspace*{-4mm}
\end{figure*}
The resulting methodology has been evaluated on the \emph{entire} Qiskit~\cite{aleksandrowiczQiskitOpensourceFramework2019} parameterized circuit library, which shows that the proposed approach can decide the equivalence of variational ansatz circuits for a high number of qubits and larger circuit depth than is even practically executable today.
The instantiation scheme, in particular, allows for checking instances that would be infeasible to handle otherwise---reducing the equivalence checking runtime by orders of magnitude in many cases.
While this method has been developed with variational algorithms in mind, the approach is much more general and works for any application that uses parameterized quantum circuits.
The implementation is open source and publicly available at \mbox{\url{https://github.com/cda-tum/qcec}}.

The remainder of this paper is structured as follows.
To keep this work self-contained, \autoref{sec:background} establishes the necessary background on quantum computing and variational quantum algorithms.
\autoref{sec:motiv-vari-quant} motivates the problem of equivalence checking of parameterized quantum circuits, discussing the shortcomings of existing equivalence checking approaches in the parameterized case and proposes a complete equivalence checking flow for parameterized circuits.
After elaborating the details of the proposed methods in \autoref{sec:an-equiv-check}, results obtained from the experimental evaluations are provided in \autoref{sec:experiments}.
Finally, \autoref{sec:conclusion} concludes this paper.

\vspace{1mm}
\section{Background}\label{sec:background}

This section briefly covers the necessary background to keep this work self-contained as well as the relevant related work from the literature. For an in-depth introduction, see~\cite{nielsenQuantumComputationQuantum2010}.

\vspace{1mm}
\subsection{Quantum Computing}\label{sec:quantum-computing}

In quantum computing, information is represented in the form of \emph{quantum bits} (short \emph{qubits}) which, contrary to the classical world, cannot only be in the basis states $\ket{0} =
\left[    \begin{smallmatrix}
  1\\0
\end{smallmatrix}\right]
$ and $\ket{1}=
\left[\begin{smallmatrix}
  0\\1
\end{smallmatrix}\right]
$, but also in a superposition
$$\alpha_0 \ket{0} + \alpha_1 \ket{1} \quad \text{with } \alpha_0, \alpha_1 \in \mathbb{C} \quad \text{and } |\alpha_0|^2 + |\alpha_1|^2 = 1.$$
Multi-qubit systems are then described as superpositions of basis states in the product Hilbert space $(\mathbb{C}^{2})^{\otimes n}$. All transformations of qubits have to be \emph{unitary transformations}, i.e., linear transformations $U: \mathbb{C}^{2^n} \rightarrow \mathbb{C}^{2^n}$ such that $U~U^\dagger = I$, where $U^\dagger$ is the conjugate transpose of $U$ and $I$ is the identity transformation. Quantum computations are typically described as \emph{quantum circuits} which are diagrams composed of wires (representing qubits) as well as boxes and interconnections on these wires called \emph{gates} (representing transformations of qubits).
If the gate set is expressive enough, any quantum computation can be written as a quantum circuit using only gates from this set.

\vspace{1mm}
\begin{example}\label{ex:basic-circuits}
  The Hadamard gate $H = \frac{1}{\sqrt{2}} \left[\begin{smallmatrix} 1 & 1 \\ 1 & -1 \end{smallmatrix}\right]$ is one of the fundamental single-qubit quantum gates because it puts a qubit in a basis state $\ket{0}$ or $\ket{1}$ into an equal superposition of these basis states. The two-qubit controlled not ($\operatorname{CNOT}$) gate
(which is natively supported on many quantum computers) flips the second qubit if the first qubit is in the $\ket{1}$ state and leaves it unchanged otherwise.
\end{example}

Although quantum algorithms are commonly described via quantum circuits across different abstraction levels, their initial description usually contains high-level gates and concepts that are not directly supported by actual quantum computers.
In fact, existing quantum computers only support a limited gate library and frequently have limited connectivity between the qubits which means that not every pair of qubits on a quantum hardware can interact in a two-qubit gate.
Hence, in order to run a quantum algorithm on an actual device, a \emph{compilation} step has to be performed which consists of synthesizing the gates of a quantum circuit to the supported gate set, mapping logical to physical qubits, inserting $\operatorname{SWAP}$ gates such that all two-qubit gates can be properly executed, and applying optimizations to reduce the size of the circuit~\cite{liTacklingQubitMapping2019,willeMappingQuantumCircuits2019,boteaComplexityQuantumCircuit2018,tanOptimalLayoutSynthesis2020}.

\begin{example}\label{ex:compilation}
  Assume the circuit in \autoref{fig:qaoa} shall be compiled to an architecture which supports a gate library consisting of $\operatorname{CNOT}$, $R_Z$, and $H$ gates, and has limited two-qubit interactions as shown on the left-hand side of \autoref{fig:qaoa-compiled}. Then, \autoref{fig:qaoa-compiled} shows a possible way of compiling this circuit.
  Because the $\operatorname{ZZ}$ gate (which is synthesized to two $\operatorname{CNOT}$ gates and an $R_Z$ gate) between physical qubit $Q_2$ and $Q_0$ cannot be executed directly, a $\operatorname{SWAP}$ gate has to be introduced into the circuit. As shown above the circuit, this $\operatorname{SWAP}$ gate is synthesized as a sequence of three $\operatorname{CNOT}$ gates. One of these $\operatorname{CNOT}$s can be cancelled with the $\operatorname{CNOT}$ of the synthesized $\operatorname{ZZ}$ gate, simplifying the circuit in the process.
\end{example}

\vspace*{-1mm}
\subsection{Variational Quantum Algorithms}\label{sec:vari-quant-algor}
\vspace*{-1mm}

Even with sophisticated compilation schemes, the noisy nature of state-of-the-art quantum computing devices makes it impossible to run far-term quantum algorithms such as Shor's algorithm~\cite{shorPolynomialtimeAlgorithmsPrime1997}, which require a substantial amount of gates. As every gate potentially introduces an error into the system and, due to the short coherence times of qubits, the maximal depth of quantum circuits is limited on NISQ devices~\cite{preskillQuantumComputingNISQ2018}.

Variational Quantum Algorithms have been proposed to allow for expressing more functionality even with low-depth circuits. The quantum computations in variational algorithms are expressed via \emph{ansatz circuits}.
These are shallow \emph{parameterized quantum circuits} $G(\theta)$ with a parameter vector $\theta = (\theta_0, \cdots, \theta_{p-1})$. Each \emph{assignment} $\sigma : \{\theta_i \mid 0 \leq i < n\} \rightarrow (-\pi, \pi]^n$ yields an \emph{instantiated circuit}~$G(\sigma(\theta))$.
The goal of a variational algorithm is to successively adapt the circuit parameters using a classical optimization routine (e.g., gradient descent) so that the resulting circuit can be used to estimate some desired quantity, i.e, the ground state of a molecule.
By using classical computing, variational ansatz circuits can be deliberately kept shallow. However, shallow circuits are generally less expressive then more complex circuits and ansatz circuits of different complexity have been developed.

\vspace*{-1mm}
\begin{example}\label{ex:variational-alg}
  The parameterized circuit shown in \autoref{fig:qaoa} represents an instance of an \emph{Quantum Alternating Operator Ansatz}~(QAOA,~\cite{hadfieldQuantumApproximateOptimization2019}) which can be used to solve problems such as quadratic unconstrained binary optimization problems.
  The name comes from the fact that a QAOA ansatz is comprised of two circuit blocks, each of which is parameterized by a different parameter.
\end{example}
\vspace*{-1mm}

Note that, throughout the iterations in the variational algorithm, only the values of the parameters change, while the general circuit structure of the ansatz stays the same.
Therefore, it is common to perform the costly compilation of an ansatz in its parametric form only once and, then, instantiate the parameters of this compiled circuit in each iteration instead of instantiating the parameters and compiling each time.
This saves a lot of overhead incurred by repeated and potentially expensive compilation steps.

  \begin{figure*}[t]
  \centering
  \resizebox{.95\linewidth}{!}{
        \begin{tikzpicture}
          \node[] (rand) {
            \begin{tikzpicture}
              \node[thick,font=\scriptsize] (G) {
                \begin{quantikz}[column sep=6pt, row sep={0.6cm,between origins}, ampersand replacement=\&]
                  \lstick{} \& \qw      \& \targ{}   \& \gate{R_Z(\theta_2)} \& \targ{} \& \qw \&      \rstick{$\cdots$}      \\
                  \lstick{} \& \gate{H} \& \ctrl{-1} \& \ctrl{1}             \& \ctrl{-1}  \& \qw \& \rstick{$\cdots$} \\
                  \lstick{} \& \gate{R_X(\theta_1)}      \& \qw       \& \targ{}              \& \gate{R_X(\theta_3)}      \& \qw \&               \rstick{$\cdots$}               \\
                \end{quantikz}
              };
              \node[thick,font=\scriptsize,below=0.2 of G] (Gp) {
                \begin{quantikz}[column sep=6pt, row sep={0.6cm,between origins}, ampersand replacement=\&]
                  \lstick{} \& \qw      \& \targ{}   \& \gate{R_Z(\theta_2)} \& \targ{} \& \qw \&      \rstick{$\cdots$}      \\
                  \lstick{} \& \gate{H} \& \ctrl{-1} \& \ctrl{1}             \& \ctrl{-1}  \& \qw \& \rstick{$\cdots$} \\
                  \lstick{} \& \qw      \& \qw       \& \targ{}              \& \gate{R_X(\theta_1 + \theta_3)}      \& \qw \&               \rstick{$\cdots$}               \\
                \end{quantikz}
              };
            \end{tikzpicture}
          };

        \node[thick,draw,font=\scriptsize,rectangle, right=0.65 of rand] (zx) {
                \begin{tikzpicture}
                  \node[] (G) {
                    \begin{ZX}[ampersand replacement=\&]
                      \zxN{} \rar \& \zxX{} \dar\rar      \& \zxZ{-\theta_2} \rar \& \zxX{} \dar\rar \& \zxN{} \rar          \& \zxN{} \rar \& \zxX{} \dar \rar \& \zxZ{\theta_2} \rar \& \zxX{} \dar \rar               \& \zxN{} \\
                      \zxN{} \rar \& \zxZ{} \rar          \& \zxZ{} \dar\rar      \& \zxZ{} \rar     \& \zxH{} \rar          \& \zxH{} \rar \& \zxZ{} \rar      \& \zxZ{} \dar\rar     \& \zxZ{}     \rar                \& \zxN{} \\
                      \zxN{} \rar \& \zxX{-\theta_3} \rar \& \zxX{} \rar          \& \zxN{} \rar     \& \zxX{-\theta_1} \rar \& \zxN{} \rar \& \zxN{} \rar      \& \zxX{} \rar         \& \zxX{\theta_2 + \theta_3} \rar \& \zxN{} \\
                    \end{ZX}
                  };

                  \node[below=.5 of G] (Gp) {
                    \begin{ZX}[ampersand replacement=\&]
                      \leftManyDots{} \zxZ[a=a1]{} \&\zxN{} \& \zxZ[a=top]{\theta_2+\theta_3+\frac{\pi}{2}} \& \zxN{}\&\zxZ[a=an]{} \rightManyDots{}               \\
                      \zxN{} \&\leftManyDots{} \zxZ[a=a2]{} \& \zxN{}  \& \zxZ[a=anm1]{} \rightManyDots{}
                      \ar[from=top,to=a1,blue,densely dotted]
                      \ar[from=top,to=a2,blue,densely dotted]
                      \ar[from=top,to=anm1,blue,densely dotted]
                      \ar[from=top,to=an,blue,densely dotted]
                    \end{ZX}
};

\node[below=.5 of Gp] (empty) {$\vdots$};

\draw[->,thick] (G) -- (Gp) node[midway,right] {Simplify};
\draw[->,thick] (Gp) -- (empty) node[midway,right] {Simplify};
                \end{tikzpicture}
        };
        \node[inner sep=0pt,above=\abovecaptionskip of zx,text width=\linewidth, align=center, font=\Large](one) {\textbf{(1) Check with \mbox{ZX-calculus}}};

        \node[thick, draw, rectangle,right=of zx] (inst){
          \begin{tikzpicture}
            \node[thick,font=\normalfont] (pick) {Pick $b$};
            \node[thick,font=\normalfont, below=.4 of pick.west,anchor=west] (assign) {Find assignment $\sigma$ s.t.};
            \node[thick,font=\normalfont, below=.1 of assign] (solve) {$A \sigma(\theta) = b$};
            \node[thick,font=\normalfont, below=1 of assign.west,anchor=west] (random) {or random $\sigma$ if last run};
          \end{tikzpicture}
        };
        \node[inner sep=0pt,above=\abovecaptionskip of inst,text width=\linewidth, align=center, font=\Large] {\textbf{(2) Instantiate}};

        \node[thick,draw,
        rectangle,right=of inst] (qcec){
          \begin{tikzpicture}
            \node[thick,font=\scriptsize,rectangle] (G) {
              \resizebox{0.25\linewidth}{!}{
                \begin{quantikz}[column sep=6pt, row sep={0.6cm,between origins}, ampersand replacement=\&]
                  \lstick{} \& \qw      \& \targ{}   \& \gate{R_Z(\sigma(\theta_2))} \& \targ{} \& \qw \&      \rstick{$\cdots$}      \\
                  \lstick{} \& \gate{H} \& \ctrl{-1} \& \ctrl{1}             \& \ctrl{-1}  \& \qw \& \rstick{$\cdots$} \\
                  \lstick{} \& \gate{R_X(\sigma(\theta_1))}      \& \qw       \& \targ{}              \& \gate{R_X(\sigma(\theta_3))}      \& \qw \&               \rstick{$\cdots$}               \\
                \end{quantikz}}
            };
            \node[thick,right=.1 of G,font=\large] (equ) {$\mathbf{\overset{?}{=}}$};
            \node[thick,font=\scriptsize,rectangle,right=0.1 of equ] (Gp) {
              \resizebox{0.25\linewidth}{!}{
                \begin{quantikz}[column sep=6pt, row sep={0.6cm,between origins}, ampersand replacement=\&]
                  \lstick{} \& \qw      \& \targ{}   \& \gate{R_Z(\sigma(\theta_2))} \& \targ{} \& \qw \&      \rstick{$\cdots$}      \\
                  \lstick{} \& \gate{H} \& \ctrl{-1} \& \ctrl{1}             \& \ctrl{-1}  \& \qw \& \rstick{$\cdots$} \\
                  \lstick{} \& \qw      \& \qw       \& \targ{}              \& \gate{R_X(\sigma(\theta_1) + \sigma(\theta_3))}      \& \qw \&               \rstick{$\cdots$}               \\
                \end{quantikz}}
            };
          \end{tikzpicture}
        };
        \node[inner sep=0pt,above=\abovecaptionskip of qcec,text width=\linewidth, align=center, font=\Large] {\textbf{(3) Check Instantiated Circuits}};

        \node[draw,thick,diamond,font=\large,inner sep = 1pt, right=of qcec.-5] (rruns) {$r$ runs?};

        \node[thick,draw,circle,fill=red!80, inner sep=1pt,minimum size=0.75cm] (x) at ($(qcec.5) + (4.75,0)$) {\color{black}{\large$\bm{\times}$}};
        \node[thick,draw,circle,fill=green!80, inner sep=1pt, minimum size=0.75cm] (y) at ($(qcec.-5) + (4.75,0)$) {\color{black}{\large$\bm{\checkmark}$}};

        \node[inner sep=1pt, align=left] (noneq) at ($(qcec.5) + (6,0)$) {Not\\equivalent};
        \node[inner sep=1pt, align=left] (equ) at ($(qcec.-5) + (6,0)$) {Equivalent};

        \draw[->,thick] (qcec.-5) -- (rruns) node[midway,above] {Yes};
        \draw[->,thick] (rruns.-90)
        .. controls +(-90:1) and +(0:1) ..($(rruns.-90)+(-1,-0.8)$) node[pos=0.1,right] {No}
        .. controls +(0:1) and +(180:1) .. ++(-12,0)
        .. controls +(180:1) and +(-90:1) .. (inst.south);

        \draw[->,thick] (rand.21) -- (zx.158);
        \draw[->,thick] (rand.-21) -- (zx.-158);

        \draw[->,thick] (zx) -- (inst) node[midway,above] {No};
        \draw[->,thick] (inst) -- (qcec) node[midway,above] {$\sigma$};

        \draw[->,thick] (rruns) -- (y) node[midway,above] {Yes};

        \draw[->,thick] (qcec.5) -- (x) node[midway,above] {No};

        \draw[->,thick, shorten >=0.1cm] (zx.-30)
        .. controls +(0:1) and +(180:1) .. ($(zx.-30)+(0.75,0)$) node[midway,above] {Yes}
        .. controls +(0:1) and +(180:1) .. ($(zx.-30)+(2.5,-1)$)
        .. controls +(0:1) and +(180:1) .. ($(zx.-30)+(18.5,-1)$)
        .. controls +(0:1) and +(-90:1) .. (y.-90);
        \end{tikzpicture}}
        \vspace*{-1mm}
  \caption{Proposed equivalence checking method for parameterized quantum circuits}
  \label{fig:check-flow}
  \vspace*{-1mm}
\end{figure*}
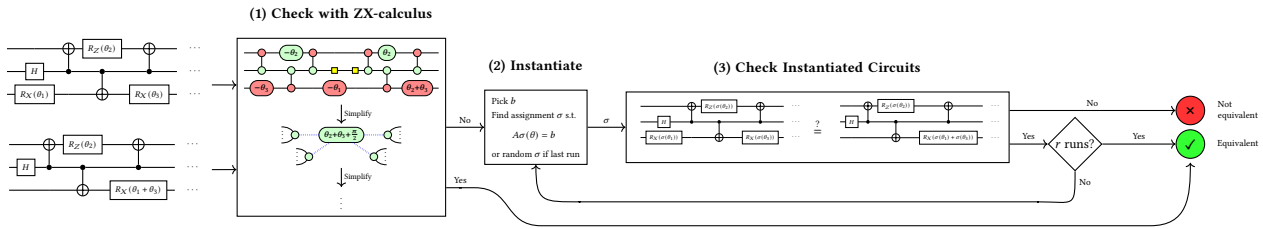

\section{Verifying Variational\\ Quantum Circuits}\label{sec:motiv-vari-quant}

Compiling a variational ansatz can significantly change its structure due to synthesized gates, $\operatorname{SWAP}$ insertions, and optimizations applied during the compilation process.
Because compilation errors are hard to detect from the results of an iteration of a variational algorithm alone, it is paramount to ensure \emph{a priori} that the compiled ansatz still adheres to its specification.
This section reviews the equivalence checking problem, describes how existing techniques can be used to solve it, and discusses why they do not work for parameterized circuits. Based on this, a methodology is proposed that deals with these shortcomings.

\vspace*{-3mm}
\subsection{Equivalence Checking}\label{sec:equivalence-checking}
\vspace*{-1mm}

In general, given two quantum circuits $G$ and $G'$ that represent the unitary matrices $U$ and $U'$, the \emph{equivalence checking problem for quantum circuits} asks whether
\[U = e^{i \gamma} U^\prime \mbox{ or, equivalently, } U^\dagger U' = e^{i \gamma}I,\]
where $\gamma \in (-\pi, \pi]$ denotes a physically unobservable global phase. In other words, equivalence checking asks whether two quantum circuits $G$ and $G'$ realize the same functionality (described by $U$ and $U'$, respectively).

Solving this problem is conceptually very simple.
After constructing the matrices $U$ and $U'$, an element-wise comparison of them shows whether the circuits are equivalent or not.
However, because of the exponential size of the matrices in question, this strategy is hardly feasible for practical quantum circuits.
This problem has already been studied extensively. As a result, more sophisticated approaches have been proposed that can frequently solve the equivalence checking problem for large circuits and many qubits~\cite{burgholzerVerifyingResultsIBM2020,amyLargescaleFunctionalVerification2019,burgholzerAdvancedEquivalenceChecking2021,kissingerReducingTcountZXcalculus2020,chun-yuAccurateBDDbasedUnitary2022}.

However, when allowing parameterized gates in the circuits to be checked, the  equivalence checking problem becomes even more general.
Checking the equivalence of two parameterized circuits $G(\theta)$ and $G'(\theta)$ requires showing that
\[
G(\sigma(\theta_0), \cdots, \sigma(\theta_{p-1}))\mbox{ and }G'(\sigma(\theta))
\]
 are equivalent for \emph{all} assignments $\sigma$.
The naive approach to circumvent this problem would be to just construct the matrices of $G(\theta)$ and $G'(\theta)$ symbolically.
But then, in addition to the exponential size of the matrices, one also has to deal with symbolic variables when constructing the matrices.
As known from computer algebra systems, trying to represent symbolic matrix entries precisely requires a lot of space for storing the coefficients of the symbolic variables.

Alternatively, one might be tempted to simply check the equivalence for one specific instantiation of two parameterized circuits to conclude equivalence. Unfortunately, this brings along a couple of difficult challenges in and of itself. On the one hand, instantiating parameters in a \emph{random} or \emph{unstructured} fashion produces circuits that are hard---if not impossible---to check with existing methods. On the other hand, instantiating parameters non-randomly can, as the following example shows, lead to false positives and, hence, also does not provide a sufficient solution.

\begin{example}\label{ex:false-positive}

  Consider the following incorrect application of a commutation rule for the $R_Z$ gate:

  \begin{center}
  \resizebox{0.8\linewidth}{!}
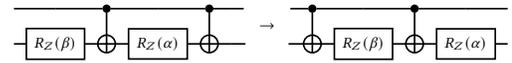
{
    \begin{quantikz}[column sep=6pt, row sep={0.6cm,between origins},ampersand replacement=\&]
      \qw \& \qw                \& \ctrl{1} \& \qw               \& \ctrl{1} \& \qw    \& \midstick[2,brackets=none]{$\rightarrow$}    \qw  \& \ctrl{1} \& \qw               \& \ctrl{1} \& \qw                \& \qw \\
      \qw \& \gate{R_Z(\beta)} \& \targ{}  \& \gate{R_Z(\alpha)} \& \targ{}  \& \qw    \&      \qw \& \targ{}  \& \gate{R_Z(\beta)} \& \targ{}  \& \gate{R_Z(\alpha)} \& \qw \\
    \end{quantikz}
  }
\end{center}

  \noindent Even though the two circuits are not equivalent for all $\alpha, \beta \in (-\pi, \pi]$, they are equivalent if $\alpha = \beta$.
  Therefore, equivalence of parametric circuits cannot be decided by checking equivalence of any one instantiation.
  \end{example}

  Because variational ansatz circuits are instantiated with different parameters in each iteration, verifying the compilation results of such circuits without symbolic equivalence checking methods requires checking the equivalence of $G(\sigma_i(\theta_0),\cdots,\sigma_i(\theta_{p-1}))$ and $G'(\sigma_i(\theta))$ for the parameter assignment $\sigma_i$ at each iteration $i$ of the hybrid optimization loop.
  Repeatedly checking equivalence in this fashion leads to obvious problems---in particular when checking a single instance is already costly.
  Therefore, dedicated methods for equivalence checking of variational ansatz circuits are desperately needed.

\vspace*{-2mm}
\subsection{Related Work}\label{sec:related-work}

Before developing solutions to the equivalence checking problem for parameterized quantum circuits completely from scratch, it is helpful to consider existing solutions for the \mbox{parameter-free} case and see whether they can be extended to the parameterized case.
Previous approaches to verification and equivalence checking broadly fall into three categories.

\begin{itemize}
\item Fully automated methods capable of checking equivalence of two specific \mbox{parameter-free} quantum circuits~\cite{chun-yuAccurateBDDbasedUnitary2022,burgholzerAdvancedEquivalenceChecking2021,amyLargescaleFunctionalVerification2019}: While these methods are able to check equivalence of many \mbox{parameter-free} circuits, they cannot be directly applied to parameterized circuits.
  More precisely, \cite{chun-yuAccurateBDDbasedUnitary2022} only works with a limited gate set as all verification rules need to be explicitly derived manually for each gate and parameter.
  The approach proposed in \cite{burgholzerAdvancedEquivalenceChecking2021} works with quantum decision diagrams which provide compact representations of unitary matrices by exploiting redundancies.
  But accurately detecting redundancies in the parametric case requires an exact representation of the respective symbolic matrix entries. Symbolic versions of decision diagrams have been employed previously~\cite{niemannOvercomingTradeoffAccuracy2020} but exhibited large space requirements due to the exploding symbolic coefficients.
  Finally, \cite{amyLargescaleFunctionalVerification2019} expresses rotations in the form of dyadic fractions similar to~\cite{niemannOvercomingTradeoffAccuracy2020}.
  But dyadic fractions are insufficient when working with gates that are not in the Clifford hierarchy as arbitrary rotations can only be approximated with dyadic fractions.

\item Semi-automatic methods using formal languages for quantum computing and proof assistants~\cite{zhouAppliedQuantumHoare2019,dhondtQuantumWeakestPreconditions2006,lewisFormalVerificationQuantum2021}:
  While these methods are very general when it comes to proving equalities in quantum computing, employing them requires expert knowledge about the languages and tools being used and, therefore, they are not suited for verification from a design automation perspective.

\item Compiler- and domain-specific methods~\mbox{\cite{taoGiallarPushbuttonVerification2022,liVerifiedCompilationQuantum2022,randReQWIREReasoningReversible2019}}:
  These methods are based on the idea of writing compiler passes in a domain-specific language.
  Those languages are constructed in a way that allows for automated verification of entire compilation flows and, therefore, are more general than an equivalence checking method that only checks the equivalence of two specific circuits.
  The drawback of these methods is that they require dedicated implementations of compilation flows which frequently have to be updated once even small aspects of the compiler are changed.
  Automated equivalence checking, on the other hand, is agnostic to future improvements in compilation and optimization methods as it only operates on the circuit and not the compilation level.
\end{itemize}

Due to these shortcomings of existing equivalence checking methods, new methods dedicated to handling parameterized methods must be developed.

\vspace*{-2mm}
\subsection{An Equivalence Checking Method for Parameterized Quantum Circuits}{\label{sec:flow}

  In this work, we propose a fully automated, efficient, and \emph{complete} method for checking the equivalence of arbitrary parameterized quantum circuits.
  To this end, a dedicated multi-stage process, as depicted in \autoref{fig:check-flow}, is used.
  The main ideas behind this method are sketched in the following.

  First, the parameterized circuits are converted into parameterized \mbox{ZX-diagram}s---a graphic description of quantum processes---that are then checked with a parametric version of the \mbox{ZX-calculus} simplification algorithm described in~\cite{kissingerReducingTcountZXcalculus2020,pehamEquivalenceCheckingParadigms2022} (step~(1) in \autoref{fig:check-flow}).
  If this check yields an affirmative answer to the question of equivalence, then the two parameterized circuits have been proven to be equivalent, and the algorithm terminates.
  However, due to the incompleteness of this rewriting approach, a non-affirmative answer does, in general, not imply that the circuits in question are \mbox{non-equivalent}.

  Therefore, if the \mbox{ZX-calculus} approach is not able to prove equivalence, further steps have to be taken.
  To this end, note that, in order to prove non-equivalence of two parameterized circuits $G(\theta)$ and $G'(\theta)$ it suffices to find \emph{one} assignment $\sigma: \theta \rightarrow (-\pi, \pi]^n$ such that $G(\sigma(\theta)) \neq G'(\sigma(\theta))$.

  As discussed above, choosing a \emph{good} assignment is a delicate issue, as random assignments may produce hard equivalence checking instances, and non-random assignments can (as discussed in \autoref{ex:false-positive}) lead to false positives.
  Instantiation can never produce false negatives, however.
  Hence, rather than choosing a random assignment, one can take advantage of the degrees of freedom provided by the parameters and instantiate the circuits in such a fashion as to make the subsequent equivalence check as simple as possible.
  In the proposed method, this is achieved by solving a linear system obtained from the expressions in parameterized gates (step~(2) in \autoref{fig:check-flow}).
  The instantiated circuits are then checked with an existing, complete equivalence checking method (step~(3) in \autoref{fig:check-flow}).
  If this method manages to prove non-equivalence of the instantiated circuits, it can be concluded that the parameterized circuits are not equivalent either.

  To counteract the possibility of false positives, multiple non-random instantiations are checked.
  In the worst case, however, all these instantiations could yield false positives.
  Therefore, the circuits are instantiated randomly as the last resort before being checked one last time.
    This has the disadvantage of handing very complicated circuits over to the equivalence checker, but the advantage---as proven in this work---is that the probability of obtaining a false positive through random instantiation is $0$.
  Hence, the output of this last check is returned as the final result.

\vspace*{-2mm}
\section{Implementation and Completeness}\label{sec:an-equiv-check}
Having the ideas and concepts discussed above (and illustrated in \autoref{fig:check-flow}), this section now provides implementation details on the respective steps and proves that the resulting overall methodology is indeed complete.

\vspace*{-2mm}
\subsection{Equivalence Checking of Parameterized Circuits with the \mbox{ZX-calculus}}\label{sec:zx-equiv}

This section describes how equivalence of parameterized circuits can be checked with the \mbox{ZX-calculus}. To this end, we first review the basics of \mbox{ZX-diagram}s and the \mbox{ZX-calculus}.

\mbox{ZX-diagram}s~\cite{vandeweteringZXcalculusWorkingQuantum2020,coeckePicturingQuantumProcesses2018} are undirected, vertex-labelled graphs consisting of two kinds of nodes (also called spiders), namely the \mbox{\emph{Z-spider}} \resizebox{!}{.7\baselineskip}{\begin{ZX}
\leftManyDots{} \zxZ[a=alpha]{\alpha}\rightManyDots{}
\end{ZX}}
 and \mbox{\emph{X-spider}}
\resizebox{!}{.7\baselineskip}{\begin{ZX}
  \leftManyDots{} \zxX[a=alpha]{\alpha}\rightManyDots{}
\end{ZX}} which carry a phase $\alpha \in [0, 2\pi)$ (a phase of $0$ is usually omitted).
Any quantum circuit can be expressed as a \mbox{ZX-diagram}.

The \mbox{ZX-calculus} is a set of axioms that can be used for equational reasoning about quantum processes solely through diagrammatic manipulations of \mbox{ZX-diagram}s.
An example of one of the axioms of this calculus is the \emph{spider fusion rule}
\resizebox{!}{\baselineskip}{    \begin{ZX}[math baseline=wantedbaseline, row sep=0.3mm,ampersand replacement=\&]
      \leftManyDots{} \zxZ[a=alpha]{\alpha}\rightManyDots{}  \& \\
      \\
      \&  \zxN{} \ar[r,3 dots] \& \zxN{}\\
\zxN[a=wantedbaseline]\\
 \& \leftManyDots{} \zxZ[a=beta]{\beta} \rightManyDots{}\\
      \ar[from=alpha, to=beta, <']
      \ar[from=beta, to=alpha, <']
    \end{ZX}}$\overset{(\mathbf{f})}{=} $ \resizebox{!}{.6\baselineskip}{\begin{ZX}[math baseline=wantedbaseline]
      \leftManyDots{} \zxZ[a=wantedbaseline]{\alpha + \beta} \rightManyDots{}
    \end{ZX}}, which allows to combine connected spiders of the same colour.

  \vspace*{1mm}
\begin{example}\label{ex:zx-encoding}
  The circuit shown in \autoref{fig:qaoa} can be translated to the a \mbox{ZX-diagram} and simplified by applying the spider fusion rule.
  \begin{center}
\resizebox{.7\linewidth}{!}
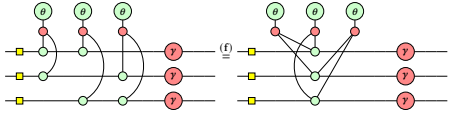
{
      \begin{ZX}[ampersand replacement=\&]
    \zxN{}      \& [\zxwCol]\zxN{}      \& [\zxwCol]\zxZ[a=phase1]{\theta}  \dar \& [\zxwCol]\zxN{}      \& [\zxwCol]\zxZ[a=phase2]{\theta} \dar \& [\zxwCol]\zxN{}      \& [\zxwCol]\zxZ[a=phase3]{\theta} \dar \& [\zxwCol]\zxN{}      \& [\zxwCol]\zxN{} \& [\zxwCol]\zxN{}             \& [\zxwCol]\zxN{}      \& [\zxwCol]\zxN{}      \& [\zxwCol]\zxN{} \\
    \zxN{}      \& [\zxwCol]\zxN{}      \& [\zxwCol]\zxX[a=gadget1]{}             \& [\zxwCol]\zxN{}      \& [\zxwCol]\zxX[a=gadget2]{}            \& [\zxwCol]\zxN{}      \& [\zxwCol]\zxX[a=gadget3]{}            \& [\zxwCol]\zxN{}      \& [\zxwCol]\zxN{} \& [\zxwCol]\zxN{}             \& [\zxwCol]\zxN{}      \& [\zxwCol]\zxN{}      \& [\zxwCol]\zxN{} \\
    \zxN{} \rar \& [\zxwCol]\zxH{} \rar \& [\zxwCol]\zxZ[a=zz11]{} \rar           \& [\zxwCol]\zxN{} \rar \& [\zxwCol]\zxZ[a=zz21]{} \rar          \& [\zxwCol]\zxN{} \rar \& [\zxwCol]\zxN{} \rar                  \& [\zxwCol]\zxN{} \rar \& [\zxwCol]\zxN{} \rar \& [\zxwCol]\zxX{\gamma} \rar \& [\zxwCol]\zxN{} \rar \& [\zxwCol]\zxN{} \rar \& [\zxwCol]\zxN{} \\
    \zxN{} \rar \& [\zxwCol]\zxH{} \rar \& [\zxwCol]\zxZ[a=zz12]{} \rar           \& [\zxwCol]\zxN{} \rar \& [\zxwCol]\zxN{} \rar                  \& [\zxwCol]\zxN{} \rar \& [\zxwCol]\zxZ[a=zz31]{} \rar          \& [\zxwCol]\zxN{} \rar \& [\zxwCol]\zxN{} \rar \& [\zxwCol]\zxX{\gamma} \rar \& [\zxwCol]\zxN{} \rar \& [\zxwCol]\zxN{} \rar \& [\zxwCol]\zxN{} \\
    \zxN{} \rar \& [\zxwCol]\zxH{} \rar \& [\zxwCol]\zxN{} \rar                   \& [\zxwCol]\zxN{} \rar \& [\zxwCol]\zxZ[a=zz22]{} \rar          \& [\zxwCol]\zxN{} \rar \& [\zxwCol]\zxZ[a=zz32]{} \rar          \& [\zxwCol]\zxN{} \rar \& [\zxwCol]\zxN{} \rar \& [\zxwCol]\zxX{\gamma} \rar \& [\zxwCol]\zxN{} \rar \& [\zxwCol]\zxN{} \rar \& [\zxwCol]\zxN{}
    \ar[from=gadget1, to=zz11]
    \ar[from=gadget1, to=zz12, o-]
    \ar[from=gadget2, to=zz21]
    \ar[from=gadget2, to=zz22, o-]
    \ar[from=gadget3, to=zz31]
    \ar[from=gadget3, to=zz32, o-]
  \end{ZX} $\overset{(\mathbf{f})}{=} $       \begin{ZX}[ampersand replacement=\&]
    \zxN{}      \& [\zxwCol]\zxN{}      \& [\zxwCol]\zxZ[a=phase1]{\theta}  \dar \& [\zxwCol]\zxN{}      \& [\zxwCol]\zxZ[a=phase2]{\theta} \dar \& [\zxwCol]\zxN{}      \& [\zxwCol]\zxZ[a=phase3]{\theta} \dar \& [\zxwCol]\zxN{}      \& [\zxwCol]\zxN{} \& [\zxwCol]\zxN{}             \& [\zxwCol]\zxN{}      \& [\zxwCol]\zxN{}      \& [\zxwCol]\zxN{} \\
    \zxN{}      \& [\zxwCol]\zxN{}      \& [\zxwCol]\zxX[a=gadget1]{}             \& [\zxwCol]\zxN{}      \& [\zxwCol]\zxX[a=gadget2]{}            \& [\zxwCol]\zxN{}      \& [\zxwCol]\zxX[a=gadget3]{}            \& [\zxwCol]\zxN{}      \& [\zxwCol]\zxN{} \& [\zxwCol]\zxN{}             \& [\zxwCol]\zxN{}      \& [\zxwCol]\zxN{}      \& [\zxwCol]\zxN{} \\
    \zxN{} \rar \& [\zxwCol]\zxH{} \rar \& [\zxwCol]\zxN[a=zz11]{} \rar           \& [\zxwCol]\zxN{} \rar \& [\zxwCol]\zxZ[a=zz1]{} \rar          \& [\zxwCol]\zxN{} \rar \& [\zxwCol]\zxN{} \rar                  \& [\zxwCol]\zxN{} \rar \& [\zxwCol]\zxN{} \rar \& [\zxwCol]\zxX{\gamma} \rar \& [\zxwCol]\zxN{} \rar \& [\zxwCol]\zxN{} \rar \& [\zxwCol]\zxN{} \\
    \zxN{} \rar \& [\zxwCol]\zxH{} \rar \& [\zxwCol]\zxN[a=zz12]{} \rar           \& [\zxwCol]\zxN{} \rar \& [\zxwCol]\zxZ[a=zz2]{} \rar                  \& [\zxwCol]\zxN{} \rar \& [\zxwCol]\zxN[a=zz31]{} \rar          \& [\zxwCol]\zxN{} \rar \& [\zxwCol]\zxN{} \rar \& [\zxwCol]\zxX{\gamma} \rar \& [\zxwCol]\zxN{} \rar \& [\zxwCol]\zxN{} \rar \& [\zxwCol]\zxN{} \\
    \zxN{} \rar \& [\zxwCol]\zxH{} \rar \& [\zxwCol]\zxN{} \rar                   \& [\zxwCol]\zxN{} \rar \& [\zxwCol]\zxZ[a=zz3]{} \rar          \& [\zxwCol]\zxN{} \rar \& [\zxwCol]\zxN[a=zz32]{} \rar          \& [\zxwCol]\zxN{} \rar \& [\zxwCol]\zxN{} \rar \& [\zxwCol]\zxX{\gamma} \rar \& [\zxwCol]\zxN{} \rar \& [\zxwCol]\zxN{} \rar \& [\zxwCol]\zxN{}
    \ar[from=gadget1, to=zz1]
    \ar[from=gadget1, to=zz2]
    \ar[from=gadget2, to=zz1]
    \ar[from=gadget2, to=zz3, -o]
    \ar[from=gadget3, to=zz2]
    \ar[from=gadget3, to=zz3]
  \end{ZX}}
  \end{center}
  Although this \mbox{ZX-diagram} looks very similar to the original circuit, the $\operatorname{ZZ}$ gates have been translated into a form (so-called phase gadgets) that has no direct interpretation as a quantum circuit anymore.
\end{example}
\vspace*{1mm}

\begin{figure}
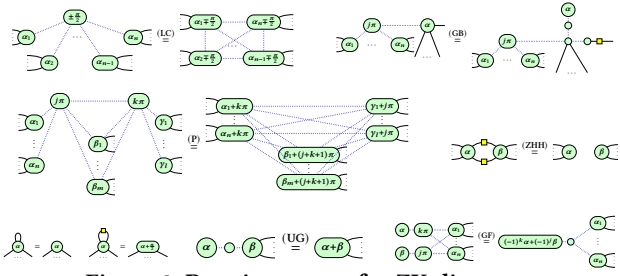

\centering
  \begin{subfigure}[c]{0.5\linewidth}
    \centering

    \begin{equation*}
      \label{eq:local_comp}
      \resizebox{0.9\linewidth}{!}{
        $				\begin{ZX}[ampersand replacement=\&]
          \zxN{} \& \zxN{}\&\zxN{} \& \zxZ[a=top]{\pm \frac{\pi}{2}} \& \zxN{}\&\zxN{} \\
          \leftManyDots{} \zxZ[a=a1]{\alpha_1} \&\zxN{} \ar[rr, 3 dots]\& \zxN{}\& \zxN{}\& \zxZ[a=an]{\alpha_n} \rightManyDots{} \\
          \\
          \zxN{} \& \leftManyDots{} \zxZ[a=a2]{\alpha_2} \& \zxN{}  \& \zxZ[a=anm1]{\alpha_{n-1}} \rightManyDots{}
          \ar[from=top,to=a1,blue,densely dotted]
          \ar[from=top,to=a2,blue,densely dotted]
          \ar[from=top,to=anm1,blue,densely dotted]
          \ar[from=top,to=an,blue,densely dotted]
        \end{ZX} \overset{(\operatorname{\mathbf{LC}})}{=}   \begin{ZX}[ampersand replacement=\&]
          \leftManyDots{} \zxZ[a=a1]{\alpha_1 \mp \frac{\pi}{2}} \&\zxN{} \& \zxN{}\& \zxN{}\& \zxZ[a=an]{\alpha_n\mp
            \frac{\pi}{2}} \rightManyDots{} \\
          \\
          \\
          \zxN{} \&  \zxN{} \&\zxN{} \ar[rr, 3 dots]\& \zxN{}\& \zxN{}\\
          \leftManyDots{} \zxZ[a=a2]{\alpha_2\mp \frac{\pi}{2}} \& \zxN{} \& \zxN{} \&  \zxN{} \& \zxZ[a=anm1]{\alpha_{n-1}\mp \frac{\pi}{2}} \rightManyDots{}
          \ar[from=a1,to=a2,blue,densely dotted]
          \ar[from=a1,to=anm1,blue,densely dotted]
          \ar[from=a1,to=an,blue,densely dotted]
          \ar[from=a2,to=anm1,blue,densely dotted]
          \ar[from=a2,to=an,blue,densely dotted]
          \ar[from=anm1,to=an,blue,densely dotted]
        \end{ZX}$}
    \end{equation*}
  \end{subfigure}
  \begin{subfigure}[c]{0.49\linewidth}
  \centering
    \begin{equation*}
      \label{eq:boundary}
      \resizebox{0.9\linewidth}{!}{
        $			\begin{ZX}[ampersand replacement=\&]
          \zxN{}\& \zxN{} \& \zxZ[a=jp]{j\pi} \& \zxN{} \& \zxN{} \& \zxN{} \& \zxZ[a=a]{\alpha}\ar[ddl] \ar[ddr] \ar[rr]\& \zxN{} \& \zxN{}\\
          \leftManyDots{} \zxZ[a=a1]{\alpha_1} \ar[rr, 3 dots] \& \zxN{} \& \zxN{} \& \zxZ[a=an]{\alpha_n} \rightManyDots{}\\
          \zxN{}\& \zxN{} \& \zxN{} \& \zxN{} \& \zxN{} \& \zxN{} \ar[rr, 3 dots]\& \zxN{} \& \zxN{} \& \zxN{}\\
          \ar[from=jp, to=a1, blue, densely dotted]
          \ar[from=jp, to=an, blue, densely dotted]
          \ar[from=jp, to=a, blue, densely dotted]
        \end{ZX} \overset{(\operatorname{\mathbf{GB}})}{=}
        \begin{ZX}[ampersand replacement=\&]
          \zxN{}                                     \& \zxN{} \& \zxN{}                                                     \& \zxN{} \& \zxN{} \& \zxN{} \& \zxN{} \& \zxZ{\alpha}\ar[d,blue,densely dotted]                                               \& [\zxwCol]\zxN{}            \& [\zxwCol] \zxN{} \& [\zxwCol] \zxN{}            \& \zxN{} \\
          \zxN{}                           \& \zxN{} \& \zxN{}                                                     \& \zxN{} \& \zxN{} \& \zxN{} \& \zxN{} \& \zxZ{}\ar[d,blue,densely dotted]                                               \& [\zxwCol]\zxN{}            \& [\zxwCol] \zxN{} \& [\zxwCol] \zxN{}            \& \zxN{} \\
          \zxN{}                                           \& \zxN{} \& \zxZ[a=v]{j\pi}\ar[dl, blue, densely dotted] \ar[dr, blue, densely dotted] \& \zxN{} \& \zxN{} \& \zxN{} \& \zxN{} \& \zxZ[a=w]{}\ar[r, blue, densely dotted] \ar[ddr]    \ar[ddl] \& [\zxwCol]\zxZ{} \ar[rr, H] \& [\zxwCol] \zxN{} \& [\zxwCol] \zxN{} \rar \& \zxN{} \\
          \leftManyDots{} \zxZ[]{\alpha_1} \ar[rr, 3 dots] \& \zxN{} \& \zxZ[]{\alpha_n} \rightManyDots{}                                                                                                                                                                                                            \\
          \zxN{}\& \zxN{} \& \zxN{} \& \zxN{} \& \zxN{} \& \zxN{} \& \zxN{} \ar[rr, 3 dots]\& \zxN{} \& \zxN{} \& \zxN{}\\
          \ar[from=v, to=w, blue, densely dotted]
        \end{ZX}$}
    \end{equation*}
  \end{subfigure}

  \begin{subfigure}[c]{0.69\linewidth}
    \centering
    \begin{equation*}
      \label{eq:pivot}
      \resizebox{0.9\linewidth}{!}{
        $			\begin{ZX}[ampersand replacement=\&]
          \zxN{} \& \zxN{}                    \& \zxZ[a=jp]{j \pi} \& \zxN{} \& \zxN{} \& \zxN{} \& \zxN{} \& \zxN{} \& \zxZ[a=kp]{k \pi} \& \zxN{} \\
          \leftManyDots{} \zxZ[a=a1]{\alpha_1}\ar[dd, 3 vdots] \& \zxN{} \& \zxN{} \& \zxN{} \& \zxN{} \& \zxN{} \& \zxN{} \& \zxN{}
          \& \zxZ[a=c1]{\gamma_1} \ar[dd, 3 vdots] \rightManyDots{}\\
          \zxN{} \& \zxN{} \& \zxN{} \& \zxN{} \& \zxN{} \& \zxZ[a=b1]{\beta_1} \rightManyDots{}\\
          \leftManyDots{} \zxZ[a=an]{\alpha_n} \& \zxN{} \& \zxN{} \& \zxN{}  \& \zxN{} \& \zxN{} \& \zxN{} \& \zxN{} \&
          \zxZ[a=cl]{\gamma_l}\rightManyDots{} \\
          \zxN{} \& \zxN{} \& \zxN{} \& \zxN{} \& \zxN{} \&\zxZ[a=bm]{\beta_m} \rightManyDots{}\\
          \ar[from=jp, to=kp, blue, densely dotted]
          \ar[from=jp, to=a1, blue, densely dotted]
          \ar[from=jp, to=an, blue, densely dotted]
          \ar[from=jp, to=b1, blue, densely dotted]
          \ar[from=jp, to=bm, blue, densely dotted]
          \ar[from=kp, to=b1, blue, densely dotted]
          \ar[from=kp, to=bm, blue, densely dotted]
          \ar[from=kp, to=c1, blue, densely dotted]
          \ar[from=kp, to=cl, blue, densely dotted]
          \ar[from=b1, to=bm, 3 vdots]
        \end{ZX}  \overset{(\operatorname{\mathbf{P}})}{=}
        \begin{ZX}[ampersand replacement=\&]
          \leftManyDots{} \zxZ[a=a1]{\alpha_1+k\pi}\ar[dd, 3 vdots] \& \zxN{} \& \zxN{} \& \zxN{} \& \zxN{} \& \zxN{} \& \zxN{} \& \zxN{}
          \& \zxZ[a=c1]{\gamma_1+j\pi} \ar[dd, 3 vdots] \rightManyDots{}\\
          \\
          \leftManyDots{} \zxZ[a=an]{\alpha_n+k\pi} \& \zxN{} \& \zxN{} \& \zxN{}  \& \zxN{} \& \zxN{} \& \zxN{} \& \zxN{} \&
          \zxZ[a=cl]{\gamma_l+j\pi}\rightManyDots{} \\
          \zxN{} \& \zxN{} \& \zxN{} \& \zxN{} \& \zxN{} \& \zxZ[a=b1]{\beta_1+(j+k+1)\pi} \rightManyDots{}\\
          \\
          \zxN{} \& \zxN{} \& \zxN{} \& \zxN{} \& \zxN{} \&\zxZ[a=bm]{\beta_m+(j+k+1)\pi} \rightManyDots{}\\
          \ar[from=a1, to=c1, blue, densely dotted]
          \ar[from=a1, to=cl, blue, densely dotted]
          \ar[from=a1, to=b1, blue, densely dotted]
          \ar[from=a1, to=bm, blue, densely dotted]
          \ar[from=an, to=c1, blue, densely dotted]
          \ar[from=an, to=cl, blue, densely dotted]
          \ar[from=an, to=b1, blue, densely dotted]
          \ar[from=an, to=bm, blue, densely dotted]
          \ar[from=b1, to=c1, blue, densely dotted]
          \ar[from=b1, to=cl, blue, densely dotted]
          \ar[from=bm, to=c1, blue, densely dotted]
          \ar[from=bm, to=cl, blue, densely dotted]
          \ar[from=b1, to=bm, 3 vdots]
        \end{ZX}  $}
    \end{equation*}
  \end{subfigure}
  \begin{subfigure}[c]{0.3\linewidth}
    \centering
    \begin{equation*}
      \label{eq:parallel}
      \resizebox{0.9\linewidth}{!}{
        $		\begin{ZX}[math baseline=base,ampersand replacement=\&]
          \leftManyDots{} \zxZ[a=base]{\alpha}\ar[rr,H,o'] \ar[rr,H,o.]  \& [\zxwCol] \zxN{} \& [\zxwCol] \zxZ{\beta} \rightManyDots{} \\
        \end{ZX} \overset{(\operatorname{\mathbf{ZHH}})}{=}
        \begin{ZX}[ampersand replacement=\&, math baseline=base]
          \leftManyDots{} \zxZ[a=base]{\alpha} \& [\zxwCol] \zxN{} \& [\zxwCol] \zxZ{\beta} \rightManyDots{} \\
        \end{ZX}$}
    \end{equation*}
  \end{subfigure}

  \begin{subfigure}[c]{0.29\linewidth}
    \centering
    \begin{equation*}
      \resizebox{0.9\linewidth}{!}{
        $		\begin{ZX}[ampersand replacement=\&,math baseline=base]
          \zxN{} \& \zxN{} \& \zxZ[a=base]{\alpha} \zxLoop{} \ar[dll] \ar[drr] \& \zxN{} \& \zxN{} \\
          \zxN{} \& \zxN{} \ar[rr, 3 dots]\& \zxN{}                                   \& \zxN{} \& \zxN{} \\
        \end{ZX} =   \begin{ZX}[ampersand replacement=\&, math baseline=base]
          \zxN{} \& \zxN{} \& \zxZ[a=base]{\alpha}  \ar[dll] \ar[drr] \& \zxN{} \& \zxN{} \\
          \zxN{} \& \zxN{} \ar[rr, 3 dots]\& \zxN{}                                   \& \zxN{} \& \zxN{} \\
        \end{ZX}
        \qquad
        \begin{ZX}[ampersand replacement=\&, math baseline=base]
          \zxN{} \& \zxN{} \& \zxZ[a=base]{\alpha} \zxLoop[90][30][H]{} \ar[dll] \ar[drr] \& \zxN{} \& \zxN{} \\
          \zxN{} \& \zxN{} \ar[rr, 3 dots]\& \zxN{}                                   \& \zxN{} \& \zxN{} \\
        \end{ZX} =
        \begin{ZX}[ampersand replacement=\&, math baseline=base]
          \zxN{} \& \zxN{} \& \zxZ[a=base]{\alpha+\frac{\pi}{2}} \ar[dll] \ar[drr] \& \zxN{} \& \zxN{} \\
          \zxN{} \& \zxN{} \ar[rr, 3 dots]\& \zxN{}                                   \& \zxN{} \& \zxN{} \\
        \end{ZX}$}
    \end{equation*}
  \end{subfigure}
  \begin{subfigure}[c]{.3\linewidth}
  \centering
    \begin{equation*}
      \label{eq:unary_gadget}
      \resizebox{0.9\linewidth}{!}{
        $					\begin{ZX}[ampersand replacement=\&]
          \zxZ{\alpha} \ar[r, blue, densely dotted] \& \zxZ{} \ar[r, blue, densely dotted] \& \zxZ{\beta} \rightManyDots{}
        \end{ZX} \overset{(\operatorname{\mathbf{UG}})}{=}
        \begin{ZX}[ampersand replacement=\&]
          \zxZ{\alpha + \beta} \rightManyDots{}
        \end{ZX}$}
    \end{equation*}
  \end{subfigure}
  \begin{subfigure}[c]{0.39\linewidth}
  \centering
    \begin{equation*}
      \label{eq:gadget_fusion}
      \resizebox{0.9\linewidth}{!}{
        $		\begin{ZX}[math baseline=base, ampersand replacement=\&]
          \zxZ[a=a]{\alpha} \ar[r,blue,densely dotted] \& \zxZ[a=ga]{k \pi} \& \zxN{} \& \zxN{} \& \zxZ[a=a1]{\alpha_1} \rightManyDots{}\\
          \zxN[a=base]{}\\
          \zxZ[a=b]{\beta} \ar[r,blue,densely dotted] \& \zxZ[a=gb]{j \pi} \& \zxN{} \& \zxN{} \& \zxZ[a=an]{\alpha_n} \rightManyDots{}\\
          \ar[from=ga, to=a1, blue, densely dotted]
          \ar[from=ga, to=an, blue, densely dotted]
          \ar[from=gb, to=a1, blue, densely dotted]
          \ar[from=gb, to=an, blue, densely dotted]
          \ar[from=a1, to=an, 3 vdots]
        \end{ZX} \overset{(\operatorname{\mathbf{GF}})}{=}
        \begin{ZX}[ampersand replacement=\&, math baseline=ga]
          \zxN{}                                       \& \zxN{} \& \zxN{} \& \zxN{} \& \zxZ[a=a1]{\alpha_1} \rightManyDots{} \\
          \zxZ{(-1)^k \alpha+ (-1)^j \beta}\ar[r,blue,densely dotted] \& \zxZ[a=ga]{}                                                     \\
          \zxN{}                                       \& \zxN{} \& \zxN{} \& \zxN{} \& \zxZ[a=an]{\alpha_n} \rightManyDots{}
          \ar[from=ga, to=a1, blue, densely dotted]
          \ar[from=ga, to=an, blue, densely dotted]
          \ar[from=a1, to=an, 3 vdots]
        \end{ZX}$}
    \end{equation*}
  \end{subfigure}\vspace*{-4mm}
  \caption{Rewrite system for \mbox{ZX-diagram}s}\label{fig:graph_like_rewriting}
  \vspace*{-4mm}
\end{figure}

The \mbox{ZX-calculus} has, among other applications, proven useful for equivalence checking~\cite{kissingerReducingTcountZXcalculus2020,pehamEquivalenceCheckingParadigms2022}. Given two quantum circuits $G$ and $G'$, one can check them for equivalence by constructing the \mbox{ZX-diagram} of $G^{-1}G'$ and, then, simplifying the diagram as much as possible using the rewriting system shown in \autoref{fig:graph_like_rewriting}.
If the \mbox{ZX-diagram} can be reduced to the identity diagram (the \mbox{ZX-diagram} consisting only of wires and no spiders), the two circuits are shown to be equivalent.

This method can be used to handle parameterized circuits as well, since all phases $\alpha_i, \beta_i, \gamma_i$ in \autoref{fig:graph_like_rewriting} are just symbolic.
For the sake of finding matches for the rewriting rules their precise value does not matter.

Whether or not the \mbox{ZX-diagram} of the circuit $G^{-1}(\theta)G'(\theta)$ can be rewritten to the identity diagram depends on the symbolic expressions appearing in the diagram.
It is reasonable to assume that all phases $\alpha_k$ are linear functions of the parameters $\theta = (\theta_0, \cdots, \theta_{p-1})$, i.e., they are of the form
$$\alpha_k = \big(\sum_{i=0}^{p-1}c_i\theta_i\big) + d \qquad c_0, \cdots, c_{p-1}, d \in \mathbb{R}.$$
This assumption is justified considering that many optimizations of quantum circuits involve gate commutations and summation of rotation angles for rotation gates~\cite{namAutomatedOptimizationLarge2018}.

Under this assumption it is easy to see that all phases resulting from the application of the rules shown in \autoref{fig:graph_like_rewriting} are also just linear functions of these parameters.
If some parameters cancel and the resulting phase has the form $k~\frac{\pi}{2}, k \in \mathbb{Z}$, new simplifications become possible.

\begin{example}\label{ex:zx-parametric}

  Cancellation of parameters during simplification can lead to further simplifications as, e.g., for following \mbox{ZX-diagram}.
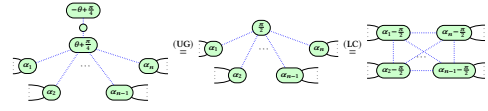
\begin{center}
  \resizebox{0.75\linewidth}{!}{
  \begin{ZX}[ampersand replacement=\&]
    \zxN{} \& \zxN{}\&\zxN{} \& \zxZ[a=phase]{-\theta+\frac{\pi}{4}} \& \zxN{}\&\zxN{}                                                                    \\
    \zxN{} \& \zxN{}\&\zxN{} \& \zxZ[a=gadget]{} \& \zxN{}\&\zxN{}                                                                    \\
    \zxN{} \& \zxN{}\&\zxN{} \& \zxZ[a=top]{\theta+\frac{\pi}{4}} \& \zxN{}\&\zxN{}                                            \\
    \leftManyDots{} \zxZ[a=a1]{\alpha_1} \&\zxN{} \ar[rr, 3 dots]\& \zxN{}\& \zxN{}\& \zxZ[a=an]{\alpha_n} \rightManyDots{} \\
    \\
    \zxN{} \& \leftManyDots{} \zxZ[a=a2]{\alpha_2} \& \zxN{}  \& \zxZ[a=anm1]{\alpha_{n-1}} \rightManyDots{}
    \ar[from=top,to=a1,blue,densely dotted]
    \ar[from=top,to=a2,blue,densely dotted]
    \ar[from=top,to=anm1,blue,densely dotted]
    \ar[from=top,to=an,blue,densely dotted]
    \ar[from=top,to=gadget,blue,densely dotted]
    \ar[from=phase,to=gadget,blue,densely dotted]
  \end{ZX}  $\overset{(\operatorname{\mathbf{UG}})}{=}$   \begin{ZX}[ampersand replacement=\&]
    \zxN{} \& \zxN{}\&\zxN{} \& \zxZ[a=top]{\frac{\pi}{2}} \& \zxN{}\&\zxN{}                                            \\
    \leftManyDots{} \zxZ[a=a1]{\alpha_1} \&\zxN{} \ar[rr, 3 dots]\& \zxN{}\& \zxN{}\& \zxZ[a=an]{\alpha_n} \rightManyDots{} \\
    \\
    \zxN{} \& \leftManyDots{} \zxZ[a=a2]{\alpha_2} \& \zxN{}  \& \zxZ[a=anm1]{\alpha_{n-1}} \rightManyDots{}
    \ar[from=top,to=a1,blue,densely dotted]
    \ar[from=top,to=a2,blue,densely dotted]
    \ar[from=top,to=anm1,blue,densely dotted]
    \ar[from=top,to=an,blue,densely dotted]
  \end{ZX}   $\overset{(\operatorname{\mathbf{LC}})}{=}$   \begin{ZX}[ampersand replacement=\&]
    \leftManyDots{} \zxZ[a=a1]{\alpha_1 - \frac{\pi}{2}} \&\zxN{} \& \zxN{}\& \zxN{}\& \zxZ[a=an]{\alpha_n-
      \frac{\pi}{2}} \rightManyDots{} \\
    \\
    \\
    \zxN{} \&  \zxN{} \&\zxN{} \ar[rr, 3 dots]\& \zxN{}\& \zxN{}\\
    \leftManyDots{} \zxZ[a=a2]{\alpha_2- \frac{\pi}{2}} \& \zxN{} \& \zxN{} \&  \zxN{} \& \zxZ[a=anm1]{\alpha_{n-1}- \frac{\pi}{2}} \rightManyDots{}
    \ar[from=a1,to=a2,blue,densely dotted]
    \ar[from=a1,to=anm1,blue,densely dotted]
    \ar[from=a1,to=an,blue,densely dotted]
    \ar[from=a2,to=anm1,blue,densely dotted]
    \ar[from=a2,to=an,blue,densely dotted]
    \ar[from=anm1,to=an,blue,densely dotted]
                                \end{ZX}}
                                \end{center}
\end{example}

If, eventually, all parameters cancel and the diagram is reduces to the identity when simplifying $G^{-1}(\theta)G'(\theta)$, it can be concluded that $G(\theta)$ is equivalent to $G'(\theta)$ entirely in parameterized form.

\vspace*{-2mm}
\subsection{Determining Instantiation Parameters}\label{sec:determ-inst-param}

Established equivalence checking methods like~\cite{burgholzerAdvancedEquivalenceChecking2021} greatly benefit from circuits that have a certain repeating structure in their functionality~\cite{pehamEquivalenceCheckingParadigms2022,burgholzerEfficientConstructionFunctional2021}.
When using such methods to check equivalence of instantiated circuits, one can exploit the degrees of freedom in the parameters to try to create such structures.
As shown in our experimental evaluations later in \autoref{sec:experiments}, this can significantly decrease the complexity of checking the resulting \mbox{parameter-free} instance.

Based on previous assumptions, if the rotation angles are of the form $\alpha_k = \big(\sum_{i=0}^{p-1}c_i\theta_i\big) + d_k$ for some constant $d_k$, then one can try to instantiate the angles $\alpha_k$ to a predefined value by solving a linear system.
This system has as many equations as there are parameterized angles in the circuits.
When choosing the angles to solve for, one has to distinguish two cases in checking \mbox{non-equivalent} circuits:

\vspace*{-1mm}
\begin{itemize}
\item The error in the circuit appears in one of the \mbox{parameter-free} gates. In this case, one can try to solve for $\alpha_k = 0$, for all parameterized angles. This effectively removes the entire parameterized gates---leading to a simpler circuit.

\item The error in the circuit appears in one of the parameterized gates. In this case, removing the gates from the circuit would mask the error and lead to false conclusions. Then, one can still try to solve for rotation angles of $0$. Nevertheless, when solving the linear system, some of the equations will, in practice, not have a solution precisely because of erroneous optimizations. The gates that cannot be removed will, therefore, usually be the ones containing an error.
\end{itemize}
\vspace*{-1mm}

In general, the linear system can be overdetermined and may not have a solution. It is NP-hard in general to find a solution such that the maximal number of equations are satisfied~\cite{guruswamiHardnessSolvingSparse2009}. One can still try to satisfy equations in a greedy fashion which, although not optimal, can be done in polynomial time.

\vspace*{-2mm}
\subsection{Completeness Proof}

If both the ZX-checker and the previously described instantiation method fail to show that two parameterized circuits are not equivalent, then it cannot be concluded with absolute certainty that the circuits are equivalent.
However, in the following, we prove that it suffices to randomly instantiate the parameters and check the equivalence of the resulting parameter-free circuits. The core observation is that, statistically, two random instantiations can be proven to \emph{almost never} produce a subsequent false positive. Here, ``almost never'' means that out of the infinitely many choices for the real-valued parameters $\theta_i$, the probability of choosing a combination of values that results in a false positive is $0$.

The following assumes a familiarity with complex analysis and measure theory to keep the proofs as concise as possible.
Any reader unfamiliar with these concepts might skip to the last paragraph of this section.

The following lemma will be the workhorse in the proof of \autoref{thm:random-complete}.

\vspace*{-3mm}
\begin{lemma}\label{lem:analytic-zero}
  Let $f: \mathbb{C}^n \rightarrow \mathbb{C}$ be an analytic function, $\lambda_n$ be the Lebesgue measure on $\mathbb{R}^n$, and $Z(f) = \{x \in \mathbb{R}^n \mid f(x)= 0\}$ the set of real zeros of $f$. If $\lambda_n(Z(f)) > 0$ then, $f = 0$.
\end{lemma}

\vspace*{-4mm}
\begin{proof}
  Since $\lambda_n(Z(f)) > 0$ implies that $Z(f)$ contains an accumulation point, this lemma follows directly from the identity principle for analytic functions.
\end{proof}
\vspace*{-4mm}

The converse of this lemma states that a non-trivial analytic function can have only countably many real zeros.
With this lemma, we can now show the desired result.

\vspace*{-3mm}
\begin{theorem}\label{thm:random-complete}
  Let $G(\theta)$ and $G'(\theta)$ be two \mbox{non-equivalent} quantum circuits with parameter vector $\theta\in(-\pi, \pi]^n$. Suppose that all rotation angles in the gates of $G(\theta)$ and $G'(\theta)$ are linear functions of~$\theta$. Then, $\mathbb{P}\{\theta \mid G(\theta) = G'(\theta)\} = 0$.
\end{theorem}
\vspace*{-4mm}

\begin{proof}
  Due to the assumption on the angles of $G(\theta)$ and $G'(\theta)$, all matrix entries of their respective gate matrices are of the form
  $e^{i \sum_{i=0}^n c_i \theta_i + d}$,
  which is a complex analytic function in $\theta$.
  The system matrices $U(\theta)$ and $U'(\theta)$ are the product of the respective gate matrices of the circuits $G$ and $G'$.
  Therefore, all entries of $U$ and $U'$ are analytic in $\theta$.
  Without loss of generality, consider the \mbox{$i,j$-th} entries $u_{i,j}(\theta), u'_{i,j}(\theta)$.
  Then, $u_{i,j}(\theta) - u'_{i,j}(\theta)$ is also analytic. By \autoref{lem:analytic-zero}, the set of zeros of $u_{i,j}(\theta) - u'_{i,j}(\theta)$ has measure zero.
  Therefore, the probability of choosing random parameters in $(-\pi, \pi]^n$ such that $u_{i,j}(\theta) = u'_{i,j}(\theta)$ is $0$.
  Since this holds for one matrix entry, it follows immediately that $\mathbb{P}\{\theta \mid G(\theta) = G'(\theta)\} = 0$.
\end{proof}
\vspace*{-3mm}

The attentive reader might argue that this fact negates the need for the previously discussed methods.
While this is true \emph{in theory}, our experimental evaluations clearly demonstrate that, in practice, equivalence checking of circuits with random rotations is a computationally difficult task. It is, therefore, only used as a last resort in the proposed equivalence checking method.

\vspace*{-1mm}
\section{Experimental Evaluation}\label{sec:experiments}

To evaluate the equivalence checking method proposed in this work, it has been implemented using a combination of C++ and Python. In order to utilize the ZX-calculus, a parameterized version of the rules in \autoref{fig:graph_like_rewriting} and the resulting equivalence checking method have been implemented. For verifying instantiated circuits the approach presented in~\cite{burgholzerAdvancedEquivalenceChecking2021} has been used. The resulting implementation is publicly available under \mbox{\url{https://github.com/cda-tum/qcec}}. The resulting tool works as a \emph{push-button} method, requiring few lines of code and no expert knowledge on verification. %
To demonstrate the effectiveness of the proposed method, we conducted a broad set of experiments discussed in the following.

\begin{table*}[t]
  \begin{minipage}[t]{0.4\linewidth}
  \centering
  \caption{Equivalent ansatz circuits}\vspace*{-4mm}
\label{tab:equivalent-benchmarks}
    \resizebox{0.9\linewidth}{!}{%
  \begin{tabular}[t]{l@{\hskip 1cm} r r r r r}
    \toprule
    Benchmark & $n$ & \#param & $|G|$ & $|G'|$ & $t$~[\si{s}]\\ \midrule\vspace{-3mm}
    \begin{filecontents*}{equivalent_benchmarks.csv}
Instance,nqubits,nParams,G,GPrime,t
TwoLocal-Circular,27,837,3295,13683,2.95
EfficientSU2-Full,27,837,12232,115099,2.11
ExcitationPreserving-FSIM-Circular,27,837,5455,9292,7.18
ExcitationPreserving-ISWAP-Circular,27,1647,12205,21913,490.59
ExcitationPreserving-ISWAP-SCA,27,567,4105,7736,13.98
ExcitationPreserving-ISWAP-Full,27,3807,49465,130977,219.36
ExcitationPreserving-FSIM-Linear,27,1607,10475,20358,9.20
ExcitationPreserving-FSIM-Linear,27,2397,15685,32577,19.11
ExcitationPreserving-ISWAP-Linear,27,1617,11785,25423,31.76
RealAmplitudes-Full,65,715,21581,226474,18.92
TwoLocal-Circular,65,1365,5331,24257,23.08
RealAmplitudes-Full,65,1365,43031,445940,31.44
ExcitationPreserving-FSIM-SCA,65,2015,13131,23878,585.24
ExcitationPreserving-FSIM-SCA,65,3965,26131,46974,1903.53
EfficientSU2-Full,65,1365,44396,446607,14.98
ExcitationPreserving-ISWAP-SCA,65,1365,9881,20770,165.96
ExcitationPreserving-ISWAP-Circular,65,455,3056,6929,11.41
TwoLocal-SCA,65,715,2731,13197,20.70
TwoLocal-Full,65,260,19046,105511,14.14
TwoLocal-SCA,127,508,1779,11093,2.41
ExcitationPreserving-FSIM-Linear,127,1264,7818,16202,8.97
RealAmplitudes-Circular,127,508,1017,8611,0.82
EfficientSU2-SCA,127,508,1525,9729,0.14
RealAmplitudes-Full,127,508,24639,287203,12.34
TwoLocal-Circular,127,508,1779,11109,1.61
RealAmplitudes-Linear,127,508,1014,7000,1.55
ExcitationPreserving-FSIM-Circular,127,1270,7875,17454,12.98
RealAmplitudes-SCA,127,508,1017,8972,0.98
ExcitationPreserving-FSIM-SCA,127,1270,7875,17316,423.05
\end{filecontents*}
\csvreader[head to column names]{equivalent_benchmarks.csv}{}{\\\Instance&\nqubits&\nParams&\G&\GPrime&\t}
    \\\bottomrule
  \end{tabular}}\\\vspace{.5mm}
\end{minipage}\hspace{0.03\linewidth}
\begin{minipage}[t]{0.563\linewidth}\centering
  \caption{Non-equivalent circuits with different instantiations}\vspace*{-4mm}
  \label{tab:non-equivalent-benchmarks}
    \resizebox{0.9\linewidth}{!}{%
  \begin{tabular}[t]{l m{0.07\linewidth} r r r r r m{0.07\linewidth} r r r m{0.07\linewidth} r r r}
    \toprule
    \multirow{2}{*}{Benchmark} & & \multirow{2}{*}{n} & \multirow{2}{*}{\#param} & \multirow{2}{*}{$|G|$} & \multirow{2}{*}{$|G'|$} & & \multicolumn{3}{c}{Flipped} & & \multicolumn{3}{c}{Shift} \\ \cmidrule(l){8-10} \cmidrule(l){12-14}
                               & &                    &                         &                      &                    &   & \#err & $t_{\text{lin}}$[\si{s}] & $t_{\text{rand}}$[\si{s}] &  & \#err & $t_{\text{lin}}$[\si{s}] & $t_{\text{rand}}$[\si{s}] \\\midrule\vspace{-3mm}
    \begin{filecontents*}{non_equivalent_benchmarks.csv}
Instance,nqubits,nParams,G,GPrime,nErrFlip,tLinFlip,tRandFlip,nErrShift,tLinShift,tRandShift
ExcitationPreserving-FSIM-Full,15,690,6061,8487,48,0.19,1178.81,9,0.20,668.49
RealAmplitudes-Full,15,60,391,1735,10,0.02,77.73,2,0.20,83.12
ExcitationPreserving-ISWAP-Full,15,375,4486,6090,24,0.21,636.63,6,0.22,542.35
TwoLocal-Full,20,40,631,1432,9,0.01,>3600,1,0.33,0.60
ExcitationPreserving-FSIM-Circular,35,560,3571,6666,21,0.15,>3600,13,0.14,>3600
ExcitationPreserving-ISWAP-Circular,35,385,2696,5557,30,0.12,>3600,6,0.12,>3600
RealAmplitudes-Full,35,210,3221,27718,117,0.16,>3600,3,0.16,>3600
EfficientSU2-Full,35,210,3431,28055,151,0.18,>3600,2,0.18,>3600
EfficientSU2-Full,35,140,2101,17705,77,0.11,>3600,1,0.10,>3600
ExcitationPreserving-FSIM-Linear,35,550,3476,4836,26,0.13,>3600,8,0.12,>3600
ExcitationPreserving-FSIM-Linear,35,344,2114,2930,8,0.10,>3600,8,0.12,>3600
TwoLocal-Full,35,70,1891,9261,44,0.16,>3600,2,0.13,>3600
TwoLocal-Full,35,105,3711,19152,92,0.43,>3600,2,0.25,>3600
TwoLocal-Full,35,140,5531,27878,128,1.21,>3600,2,0.43,>3600
ExcitationPreserving-ISWAP-Full,35,665,8436,19992,93,2.03,>3600,11,2.15,>3600
ExcitationPreserving-FSIM-Full,35,1260,11411,25245,124,2.03,>3600,35,11.40,>3600
ExcitationPreserving-ISWAP-Full,35,1295,16801,41715,211,7.79,>3600,26,204.66,>3600
EfficientSU2-Full,40,160,2701,19823,91,0.28,>3600,2,0.28,>3600
TwoLocal-Full,40,160,7221,37293,192,2.31,>3600,1,0.73,>3600
ExcitationPreserving-FSIM-Circular,40,640,4081,5681,29,0.16,>3600,10,0.16,>3600
ExcitationPreserving-ISWAP-Linear,40,435,3011,4001,24,0.15,>3600,2,0.12,>3600
ExcitationPreserving-FSIM-Linear,40,630,3986,5546,21,0.18,>3600,12,0.15,>3600
RealAmplitudes-Full,40,160,2541,22750,95,0.25,>3600,2,0.21,>3600
ExcitationPreserving-FSIM-Linear,45,710,4496,6256,40,0.22,>3600,14,0.18,>3600
ExcitationPreserving-FSIM-Circular,45,450,2791,5827,37,0.15,>3600,13,0.14,>3600
ExcitationPreserving-ISWAP-Linear,45,490,3396,4511,23,0.19,>3600,4,0.16,>3600
ExcitationPreserving-ISWAP-Circular,45,495,3466,7921,40,0.24,>3600,11,0.20,>3600
ExcitationPreserving-FSIM-Circular,45,720,4591,9010,41,0.23,>3600,10,0.21,>3600
ExcitationPreserving-ISWAP-Circular,45,315,2116,4975,23,0.14,>3600,10,0.14,>3600
EfficientSU2-Circular,45,270,811,4330,19,0.13,>3600,6,0.10,>3600
ExcitationPreserving-FSIM-Linear,50,494,3044,4220,28,0.16,>3600,14,0.14,>3600
\end{filecontents*}
\csvreader[head to column names]{non_equivalent_benchmarks.csv}{}{\\\Instance&&\nqubits&\nParams&\G&\GPrime&&\nErrFlip&\tLinFlip&\tRandFlip& &\nErrShift&\tLinShift&\tRandShift                                                                                                                                                            }    \\\bottomrule
  \end{tabular}}
\end{minipage}
\\
{\tiny $n$: Number of qubits \hspace*{0.40cm} \#param: Number of distinct parameters \hspace*{0.40cm} \emph{$\vert G \vert$}: Gate count of uncompiled ansatz \hspace*{0.40cm} \emph{$\vert G^\prime \vert$}: Gate count of compiled ansatz \\\vspace{-1.5mm} $t$: Runtime of ZX-checker \hspace*{0.40cm} \#err: Number of errors \hspace*{0.40cm}\vspace*{-.8\baselineskip} $t_\text{\tiny lin}$: Runtime with linear system instantiation \hspace*{0.40cm} $t_\text{\tiny rand}$ Runtime with random instantiation}\vspace*{-3mm}
\end{table*}

\vspace*{-2mm}
\subsection{Setup}\label{sec:setup}

To test a wide range of parameterized ansatz circuits, the proposed method was evaluated on the \emph{entire} available library of parameterized ansatz circuits provided by the Qiskit circuit library~\cite{aleksandrowiczQiskitOpensourceFramework2019}. The circuits have been created with varying depths and different entanglement patterns (linear, circular, full, and SCA).

In order to also evaluate the proposed method on \mbox{non-equivalent} circuits, two kinds of errors were injected into the compiled circuits. The first type of error were incorrect applications of control and target qubits in two-qubit gates. The goal of the first type of error was to mimic situations where the control and targets of certain \emph{two-qubit} gates have been erroneously exchanged---a common mistake that can happen in compilers.
To this end, every two-qubit gate had a low chance (0.5\%) of being flipped.
The second type of error was concerned with errors in the parameters.
Many optimizations on rotation gates involve phase shifts in the rotation angles of a gate. Therefore, with a low probability (1\%), each parameterized angle had a chance of having an erroneous phase shift applied to it.

All circuits have been compiled using \emph{qiskit-terra 0.21.0} with optimization level \emph{O2}.
The target devices for compilation were the $20$-qubit \emph{ibmq-singapore}, the $27$-qubit \emph{ibmq-cairo}, the $65$-qubit \emph{ibmq-manhattan} and the $127$-qubit \emph{ibmq-washington} devices.
The circuits have been compiled to the next smallest viable architecture.

\vspace*{-2mm}
\subsection{Results and Discussion}\label{sec:discussion}

\autoref{tab:equivalent-benchmarks} shows the runtime ($t$) of the proposed approach for circuits with varying numbers of qubits ($n$), parameters ($\#param$), and gates for both the original ($G$) and compiled version ($G'$) of each \emph{equivalent} ansatz circuit~\footnote{Due to space restrictions, only a representative sample of the results is provided. All results are available at \mbox{\url{https://github.com/cda-tum/qcec}}}.
All considered circuits were successfully verified completely with the ZX-calculus, and no instantiation was necessary.
As can be seen from these results, the proposed approach scales to the largest quantum systems available today and is capable of verifying the compilation results of any of the variational algorithms that are currently being explored for near-term applications.

\autoref{tab:non-equivalent-benchmarks} compares the instantiation approach proposed in \autoref{sec:determ-inst-param} with random instantiation for both considered types of errors.
The table shows runtimes for both the proposed instantiation method using linear systems ($t_\text{lin}$) and random instantiation ($t_\text{rand}$).
These are listed along with the number of errors ($\#err$) for flipped two-qubit gate errors (Flipped) and phase shift errors (Shift).
While the state-of-the-art method from~\cite{burgholzerAdvancedEquivalenceChecking2021} used to conduct the equivalence check of the instantiated circuits quickly runs into the set timeout of $1$\si{h} when using random instantiation, the instantiation approach through solving linear systems allows to conclude the non-equivalence for a much larger range of circuits---in many cases within fractions of a second. This underlines the point made at the end of the previous section that, while random instantiations would be ``good enough'' for equivalence checking parameterized circuits in theory, they are hardly practical.

\vspace{-1mm}
\section{Conclusion}\label{sec:conclusion}
\vspace*{-1mm}

This paper proposes a methodology that, for the first time, allows for checking the equivalence of parameterized quantum circuits.
The proposed equivalence checking methodology is complete, i.e., it can check equivalence for any pair of quantum circuits.
This is achieved by combining a \mbox{ZX-calculus} approach working directly on parameterized circuits and an instantiation strategy to create \mbox{parameter-free} circuits that are efficiently checkable by existing equivalence checking methods.
The resulting tool has been demonstrated to be capable of verifying the compilation of any of the variational ansatz circuits currently being explored for near-term quantum applications---even for the largest available quantum architectures.

\vspace*{-2mm}
\subsection*{Acknowledgements}
\vspace*{-1mm}
This work received funding from the European Research Council (ERC) under the European Union’s Horizon 2020 research and innovation program (grant agreement No. $101001318$), was part of the Munich Quantum Valley, which is supported by the Bavarian state government with funds from the Hightech Agenda Bayern Plus, and has been supported by the BMWK on the basis of a decision by the German Bundestag through project QuaST.

\vspace*{-1mm}
\printbibliography
\end{document}